\documentclass[aip,reprint,superscriptaddress]{revtex4-1}  
\usepackage{graphicx}  
\usepackage{dcolumn}   
\usepackage{bm}        
\usepackage{amssymb}   
\graphicspath{{Pictures/}} 
\usepackage{booktabs}
\usepackage{amsmath,amssymb,color,float,setspace}
\usepackage[utf8]{inputenc}
\usepackage[T1]{fontenc}
\usepackage{titlesec}
\usepackage{booktabs}
\usepackage{booktabs}
\usepackage[bottom]{footmisc}
\usepackage{multirow}

\usepackage{hhline}

\begin{document}

\title{Superconducting and Antiferromagnetic Properties of Dual-Phase V$_3$Ga}
\author{Michelle E. Jamer}
\affiliation{Physics Department, United States Naval Academy, Annapolis, MD 20899, USA}
\author{Brandon Wilfong}
\affiliation{Physics Department, United States Naval Academy, Annapolis, MD 20899, USA}
\author{Vasiliy D. Buchelnikov}
\affiliation{Faculty of Physics, Chelyabinsk State University, 454001 Chelyabinsk, Russia}
\affiliation{National University of Science and Technology "MISiS", 119049 Moscow, Russia}
\author{Vladimir V. Sokolovskiy}
\affiliation{Faculty of Physics, Chelyabinsk State University, 454001 Chelyabinsk, Russia}
\affiliation{National University of Science and Technology "MISiS", 119049 Moscow, Russia}
\author{Olga N. Miroshkina}
\affiliation{Faculty of Physics, Chelyabinsk State University, 454001 Chelyabinsk, Russia}
\affiliation{Department of Physics, School of Engineering Science, LUT University, FI-53851 Lappeenranta, Finland}
\author{Mikhail A. Zagrebin}
\affiliation{Faculty of Physics, Chelyabinsk State University, 454001 Chelyabinsk, Russia}
\affiliation{National Research South Ural State University, 454080 Chelyabinsk, Russia}
\author{Danil R. Baigutlin}
\affiliation{Faculty of Physics, Chelyabinsk State University, 454001 Chelyabinsk, Russia}
\affiliation{Department of Physics, School of Engineering Science, LUT University, FI-53851 Lappeenranta, Finland}
\author{Jared Naphy}
\affiliation{Physics Department, United States Naval Academy, Annapolis, MD 20899, USA}
\author{Badih A. Assaf}
\affiliation{Physics Department, University of Notre Dame, South Bend, IN 46556}
\author{Laura H. Lewis}
\affiliation{Chemical Engineering Department, Northeastern University, Boston, MA}
\author{Aki Pulkkinen}
\affiliation{Department of Physics, School of Engineering Science, LUT University, FI-53851 Lappeenranta, Finland}
\author{Bernardo Barbiellini}
\affiliation{Department of Physics, School of Engineering Science, LUT University, FI-53851 Lappeenranta, Finland}
\affiliation{Physics Department, Northeastern University, Boston, MA}
\author{Arun Bansil}
\affiliation{Physics Department, Northeastern University, Boston, MA}
\author{Don Heiman}
\affiliation{Physics Department, Northeastern University, Boston, MA}
\date{\today}

\begin{abstract}
The binary compound V$_3$Ga can exhibit two near-equilibrium phases, consisting of the A15 structure that is superconducting, and the Heusler D0$_3$ structure that is semiconducting and antiferromagnetic. Density functional theory calculations show that the two phases are closely degenerate, being separated by only $\pm$10~meV/atom. Magnetization measurements on bulk-grown samples show superconducting behavior below 14~K. These results indicate the possibility of using V$_3$Ga for quantum technology devices utilizing both superconductivity and antiferromagnetism at the same temperature.  
\end{abstract}

\maketitle
\section{Introduction}
\begin{figure}
\centering
\includegraphics[width=0.4\textwidth]{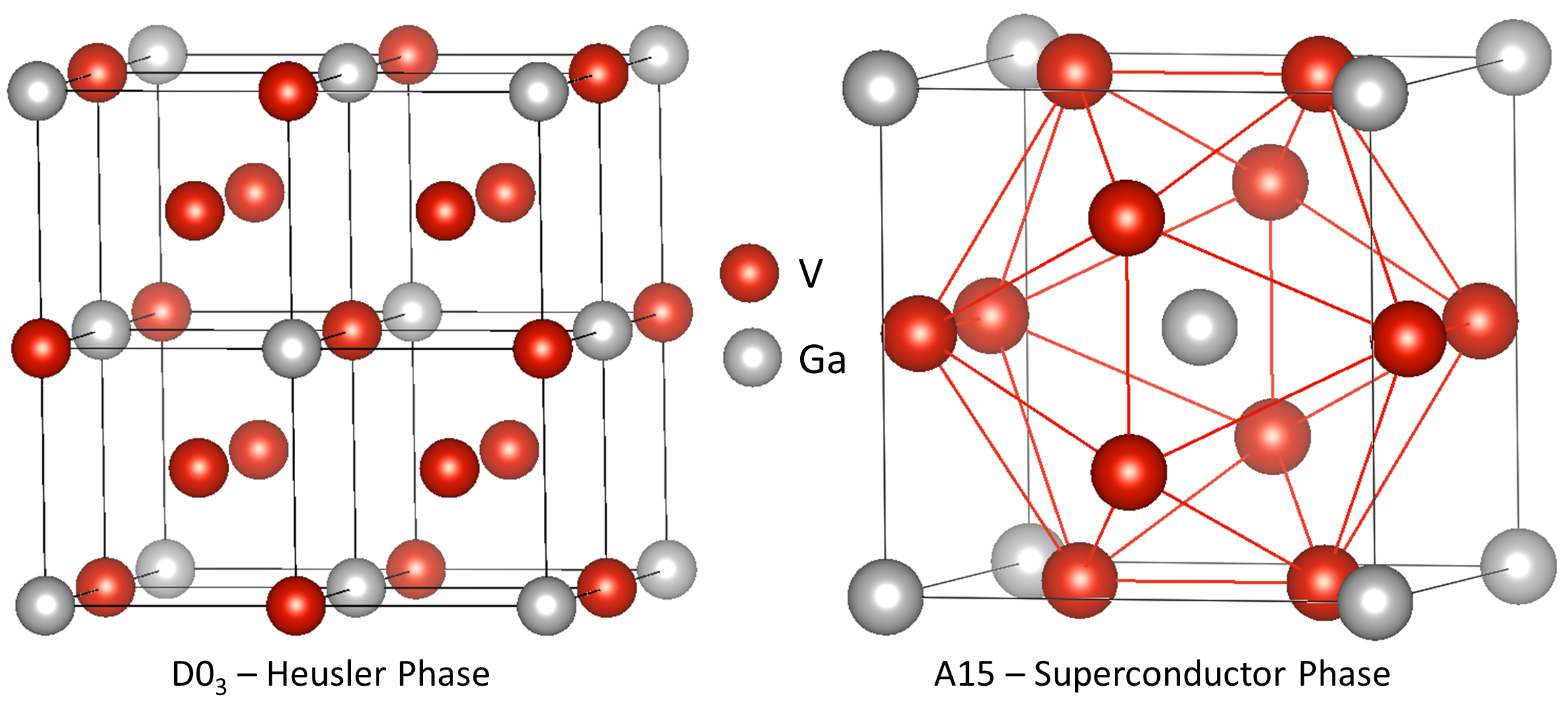}
\caption{\small D0$_3$ (space group Fm$\bar3$m no. 225, prototype BiF$_3$) and $\beta$-W A15 (space group Pm$\bar3$n no. 223, prototype Cr$_3$Si) crystal structures of V$_3$Ga.}
\label{fig1}
\end{figure}

Combining superconductivity and antiferromagnetism could find useful applications in quantum spintronic devices. Superconductivity and magnetism were once thought to be mutually exclusive, as magnetic fields are efficient at closing the superconducting gap. Nevertheless, it was found that superconducting materials could contain 3$d$  magnetic transition metal atoms and magnetic lattices as well \cite{Canfield-1998}. Following that, high-$T_c$ cuprate superconductors were found to have exceedingly strong magnetic exchange\cite{RevModPhys.78.17}, while superconducting Fe-pnitides were found to have large Fe moments of several Bohr magnetons\cite{Mizuguchi,Lumsden_2010}.
Of interest here are binary vanadium compounds such as V$_3$Al that belong to a class of simple superconductor materials with A15 crystal structure \cite{Du_2013,Bansil1999,Testardi-1997,Ohshima_1989}. Interestingly, V$_3$Al was also synthesized in a non-superconducting D0$_3$ Heusler phase with antiferromagnetic (AFM) order \cite{Jamer4}. This D0$_3$ phase of V$_3$Al was predicted to be a gapless semiconductor\cite{Gao2013,V3AlGalanakis}, and was found experimentally\cite{Jamer4} to be a G-type antiferromagnet having a Ne\'{e}l temperature of $T_N = $~600~K. One can draw the conclusion that V$_3$Z-type compounds represent a class of hybrid materials with superconducting and magnetic properties at the same temperature, which could have applications in possible fault-tolerant quantum computers hosting Majorana modes and in other important quantum technology applications.

Another well-known compound in this family is V$_3$Ga, which has been used in superconducting applications for many years \cite{markiewicz_1977}. This material has been known since 1956 to have remarkable low-temperature properties related to elastic constants, Knight shifts, electrical resistance, magnetic susceptibility, superconductivity and it has been investigated extensively both experimentally and theoretically  (see e.g. Refs.~\cite{Matthias-1956,Weger-1964, Izyumov_Kurmaev, Testardi-1975, Klein-1978, Jarlborg-1979,MaterialsProj}). The critical temperature of superconducting V$_3$Ga in the A15 phase is 14~K. 

V$_3$Ga can exist in two near-equilibrium phases, the A15 superconducting phase and the AFM D0$_3$ phase - an interesting and potentially useful result of their similar formation energies. Since the arrangement of atoms in binary V$_3$Ga can accommodate both D0$_3$ and A15 structures, shown in Fig.~\ref{fig1}, one must study the stability of these two phases. Density functional theory~(DFT) was used here to compute the formation energies as a function of crystalline and magnetic structures. Previous theoretical calculations for the D0$_3$ structure of V$_3$Ga by Galanakis {\it{et al.}} \cite{V3AlGalanakis} have predicted a Heusler G-type AFM phase with a Ne\'{e}l temperature well above room temperature, which makes the compound attractive for applications~\cite{Jamer4,JamerCrCoGa,JamerPRA2017}. A recent study also reported on the similar formation energies of the two structures.\cite{He-2019} The present magnetization measurements on bulk samples show a strong Meissner effect indicating a superconducting transition temperature of 14~K.

\section{Experimental Details and Results}
Bulk samples of V$_3$Ga were synthesized via arc-melting using an Edmund Buehler MAM-1. The ingots were subsequently annealed at 1050$^\circ$C for 48 hours in an Argon environment to promote homogeneity and quenched in an ice bath. The composition was confirmed using energy dispersive spectroscopy (EDS) to be within $\pm$2\% of the nominal composition. Magnetic measurements were performed in a Quantum Design MPMS XL-5 SQUID magnetometer in magnetic fields up to 5~T and temperatures from T~=~2 to 400~K. Synchrotron X-ray diffraction was done using beamline 11-BM at the Advanced Photon Source (APS). The structure was refined using TOPAS.\cite{TOPAS} Overall, the structural results determined that there was a mixed phase of both A15 and D0$_3$ of V$_3$Ga. The Rietveld refinement analysis determined that the A15 was the predominant phase at 81\% and the D0$_3$ accounted for 19\% of the structural analysis (Supplemental Information).

\begin{figure}
	\centering
	\includegraphics[width=0.4\textwidth]{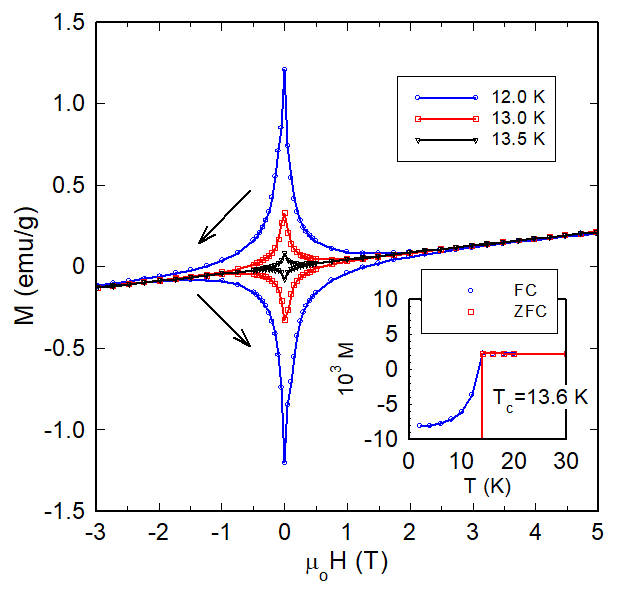}
	\caption{\small Magnetization of superconducting V$_3$Ga versus magnetic field at low temperatures under Zero field cooling (ZFC), showing a hysteretic peak around H~=~0 characteristic of flux pinning in a type-II superconductor. The inset shows the temperature-dependent Meissner flux exclusion  below $T_C=13.6$~K taken in a field of H~=~500~Oe.}
	\label{fig2}
\end{figure}

\begin{figure}
	\centering
	\includegraphics[width=0.38\textwidth]{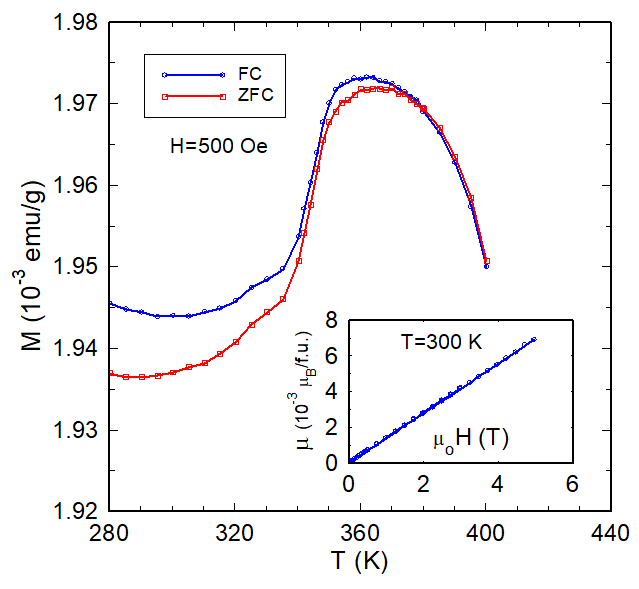}
	\caption{\small Magnetic properties of AFM V$_3$Ga versus temperature and magnetic field. Magnetization versus temperature taken at H~=~500~Oe shows a peak at 360~K, indicating a magnetic transition. (inset) Magnetic moment versus magnetic field taken at $T$~=~300~K, showing a small moment of only $\mu$~=~0.007~$\mu_B$/f.u. at $\mu_0$H~=~5~T.}
	\label{fig3}
\end{figure}

 Results of the magnetic measurements are shown in Figs.~\ref{fig2} and \ref{fig3}. In particular, Fig.~\ref{fig2} shows a magnetic hysteresis from the Meissner effect at low temperature, which is characteristic of flux pinning of the A15 type-II superconducting phase. The superconducting transition temperature $T_c=13.6$~K is deduced from the plot of the magnetization as a function of temperature as shown in the inset of Fig.~\ref{fig2}. Dimensionless magnetic susceptibility is shown in Supplemental Information which indicates $\sim$ 90\% superconducting volume fraction in line with expectations from the compositional analysis of the X-ray diffraction data through Rietveld refinement. The inset of Fig.~\ref{fig3} shows the magnetic moment as a function of applied field at $T$~=~300~K. The moment is linear up to at least $\mu_0$H~=~5~T, where the moment is only $\mu~=~0.007~\mu_B$ per formula unit (f.u.). This linear-in-H moment is similar to that found in the AFM D0$_3$ phase of V$_3$Al\cite{Jamer4}. A focus on the superconducting properties of the A15 phase is presented in Supplemental Information which shows an H$_{c2}$ of about 3.5 T for isothermal magnetization at 10 K. This value is significantly lower than the previously accepted value of 15 T at 10 K for the A15 V$_{3}$Ga phase. \cite{Foner, Decker} The decrease in H$_{c2}$ is most likely due to the antiferromagnetic phase from the D0$_3$ component which attributes significant magnetic signal at higher field. Previous reports of off-stoichiometry A15 V$_{3}$Ga \cite{Foner} did not alter the H$_{c2}$ as drastically as the observed in the current system, therefore it is reasonable to attribute this significant decrease in H$_{c2}$ as due to the antiferromagnetic signal of another magnetic V$_{3}$Ga phase present in the as-synthesized compound mixture. At higher temperatures there is a notable peak in the temperature dependence of the low-field (H~=~500~Oe) moment at $T$~=~360~K, shown in Fig.~\ref{fig3}. A similar feature was seen in the temperature-dependent resistivity of V$_3$Ga at 350~K\cite{He-2019}, however, the peak was not assigned to a magnetic transition. The similar temperature-dependent features in both the magnetization and the resistivity could arise from either a structural or a magnetic transition. A small peak in the magnetization of an AFM is generally characteristic of a Ne\'{e}l temperature\cite{Jamer4}, so there is a reasonable case to assign the small peak observed here in the magnetization of V$_3$Ga to some type of magnetic transition. 

\begin{table*}[tbp]
	\caption{The calculated equilibrium magnetic states (\textit{Mag. st.}), lattice constants ($a_0$ in \AA), atom-resolved and total magnetic moments ($\mu_{V_i}$ and $\mu_{tot}$ in $\mu_B$/f.u.), total energy ($E_0$ in eV/atom), as well as the energy difference between the D$0_3$ and A15 structures ($\Delta E_{D0_3-A15}$ in eV/atom) for V$_3$Ga. The results are shown for various exchange-correlation approaches described in the text. For the A15 structure, the SCAN solutions for FM and AFM-III are almost degenerate and within 5 meV/atom.}
	\label{table-1}
		\begin{tabular}{|l|ccccccc|ccccccc|c|}
		\hline
			\multicolumn{1}{|c}{} & \multicolumn{7}{|c|}{D0$_3$}                                                  & \multicolumn{7}{c|}{A15}                                                 & \multirow{2}{*}{$\Delta E$} \\
			& \textit{Mag. st.} & $a_0$ & $\mu_{V_1}$    & $\mu_{V_2}$   & $\mu_{V_3}$   & $\mu_{tot}$ & $E_0$  & \textit{Mag. st.} & $a_0$ & $\mu_{V_1}$    & $\mu_{V_2}$   & $\mu_{V_3}$   & $\mu_{tot}$ & $E_0$  &                         \\ \hline
			LDA &  AFM     &     5.902     &   0.429     &  -0.429     & 0.0       &      0.0 &    -8.487                   & FM    & 4.678	&0.089&	0.164&	0.112&	0.368&	-8.572&     0.085                   \\
			GGA                  & AFM   &  6.064  &	-1.314 &	1.314 &	0.0 &	0.0 &	-7.600
			  & FM  & 4.788 &	0.222 &	0.390 &	0.303 &	0.916 &	-7.645 & 0.045
			                      \\
			GGA+$U$                & AFM   &    6.130  &	-1.916 &	1.917 &	0.0 &	0.0 &	-6.647 
			 & AFM-III   &   4.879  &	$\pm$1.351  & 	$\pm$1.508 &	$\pm$1.502 &	0.0 &	-6.632 & -0.015
			                      \\
			SCAN                 & AFM   & 6.035&	-1.848&	1.848&	0.0&	0.0&	-17.486
			    &FM  & 4.744 &	0.308 &	0.523 &	0.414 &	1.245 &	-17.442 & -0.044 \\ 
			  & & & & & & & & AFM-III&	4.744	& $\pm$0.268 &	$\pm$0.326 &	$\pm$0.330 &	0.0 &	-17.437 & - \\
   \hline                 
		\end{tabular}%
\end{table*}

\section{Computational Details and Results}

Density Functional Theory (DFT) within the projector augmented wave~(PAW) as implemented in  VASP~\cite{Kresse-1996,paw} was used for the electronic structure calculations. Various approximations were considered for the exchange-correlation~(XC) energy, such as the local density approximation~(LDA)~\cite{Perdew_LDA}, the generalized gradient approximation~(GGA)~\cite{Perdew-1991,Burke-1997,Perdew1996}, GGA+$U$ (with the Hubbard $U$ correction), and the strongly constrained and appropriately normed~(SCAN)~meta-GGA~\cite{Perdew-1999,Tao-2003,Sun-2015}. We have allowed full lattice and spin relaxation. The calculations were converged to an accuracy of 10$^{-8}$~eV,  while a convergence criterion in the optimization for the residual forces was 10$^{-7}$~eV/\AA. Concerning the GGA+$U$, the Coulomb integrals $U$ = 2.0 eV and $J$ = 0.67 eV have been used that were provided by He \textit{et al.}~\cite{He-2019}.
 
 The main results are summarized in Table~\ref{table-1}, while more details are contained in the Supplemental Materials. The general trend of correlation effects beyond GGA is to stabilize the D0$_3$ solution with respect to the A15 one. The ground state (GS) in both LDA and GGA is the ferromagnetic (FM) phase with A15 structure. When correlation effects are included, the GS becomes the AFM G-type D0$_3$ solution. 
 Regarding the A15 solution, the effect of correlations is to stabilize an AFM-III solution~\cite{He-2019,Supplemental} with respect to FM one. Within SCAN, the FM and AFM-III solutions in the A15 structure are almost degenerate and within 5 meV/atom (FM is marginally more stable). However, GGA+$U$ fully stabilizes the AFM-III solution. The AFM-III solution having a net zero magnetic moment is more compatible with the superconducting properties of A15 structure.
 
\begin{figure}
	\centering
	\includegraphics[width=\columnwidth]{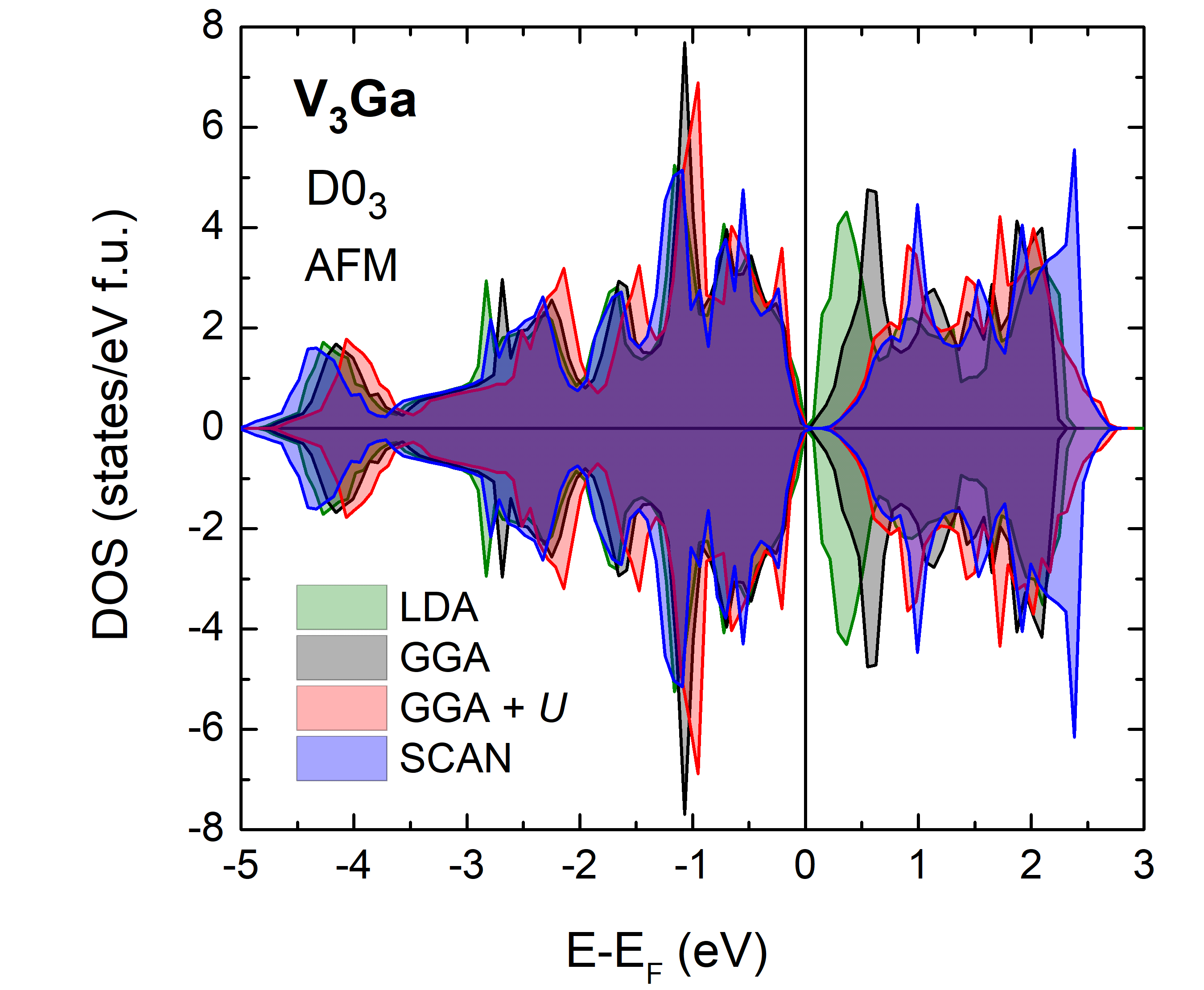}
	\vfill
	\includegraphics[width=\columnwidth]{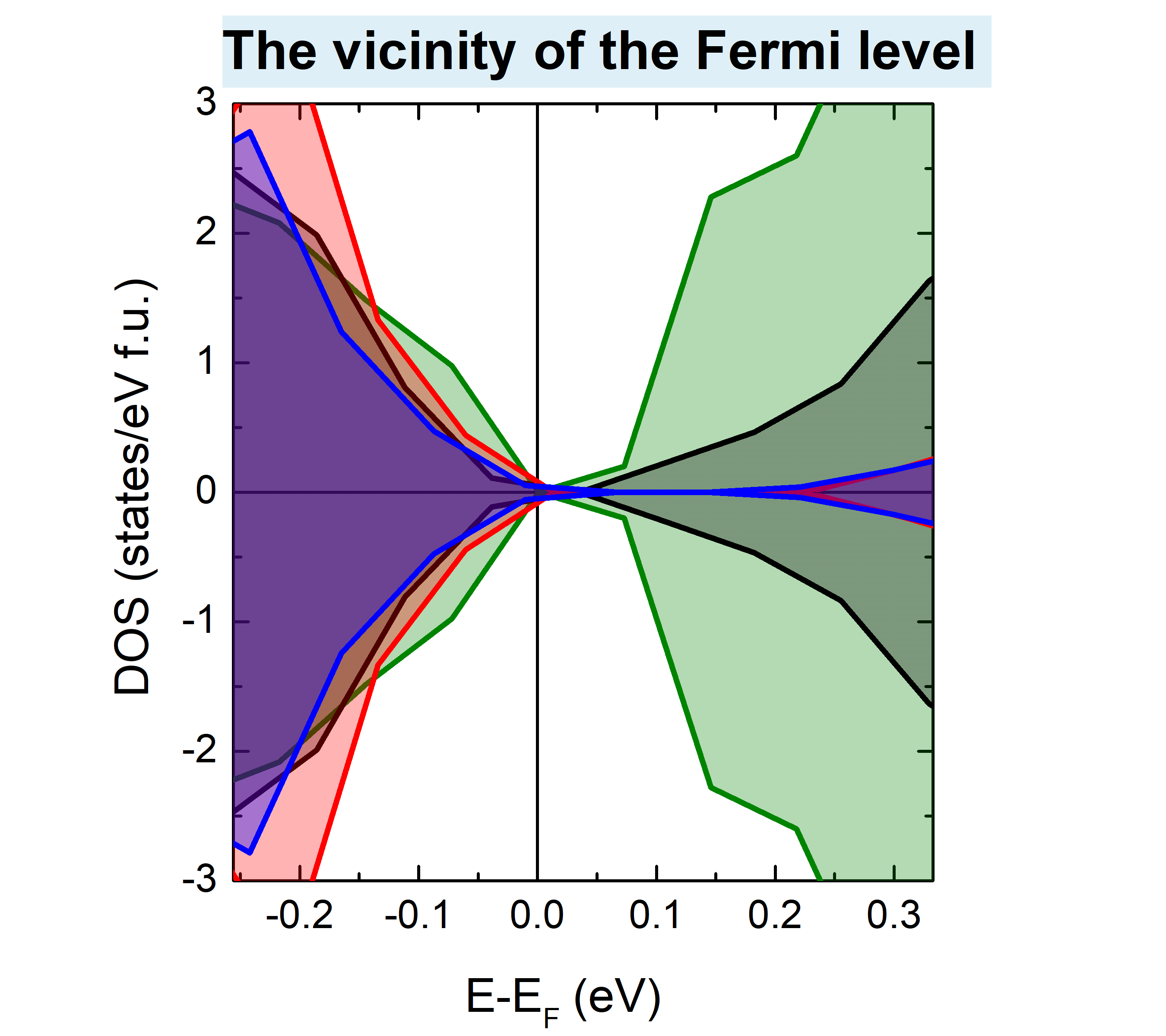}
	\caption{\small Total DOS for AFM D0$_3$ structure calculated with LDA, GGA, GGA+$U$ and SCAN methods. The lower figure shows the gapless region near the Fermi level on an expanded energy scale.}
    \label{fig4}
\end{figure}

 Given the observation of the dual phase in the present V$_3$Ga samples, SCAN may exaggerate the stabilization of the D0$_3$ solution, while GGA+$U$ gives almost degenerate A15 and D0$_3$ solutions within 15 meV/atom, a value found previously in DFT solutions\cite{He-2019} and is in better accord with the experiment.
SCAN is also known to exaggerate the magnetic moment of transition metal atoms, which are well described within GGA~\cite{isaacs2018,fu2018,ekholm2018}. In order to alleviate this problem, a modification of SCAN with deorbitalization has been suggested recently~\cite{deorbit}. 

  The calculated electronic structure is represented by density of states (DOS) shown in Fig.~\ref{fig4}. LDA and GGA schemes give an almost gapless semiconductor, while the effects of correlation beyond GGA within GGA+$U$ and SCAN lead to the opening of gap with size of about 0.2~eV. Similar gap opening has been observed by Buchelnikov~\textit{et~al.}~\cite{Buchelnikov-2019} in other Heusler alloys.
  
\begin{figure}
	\centering
	\includegraphics[width=0.5\textwidth]{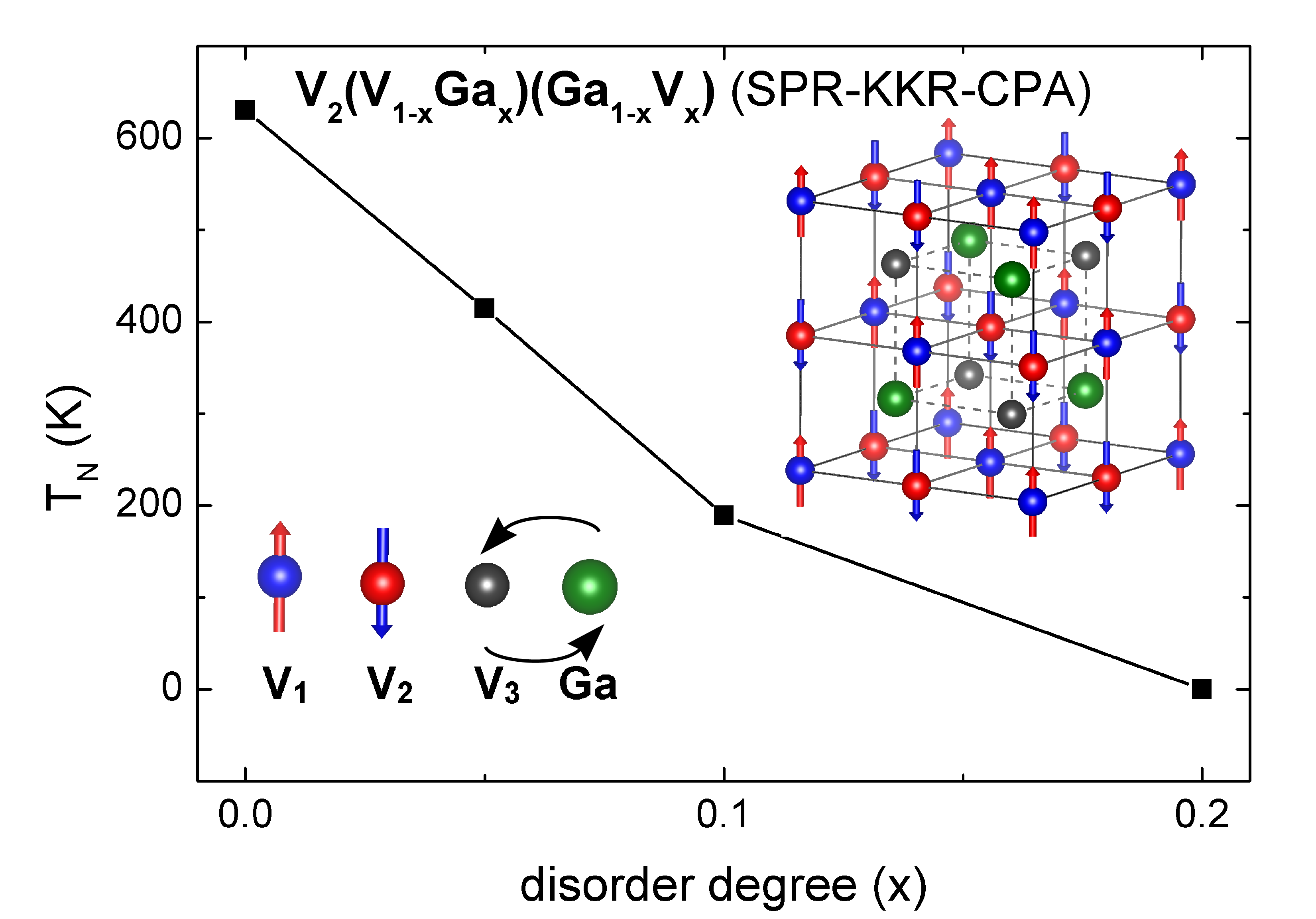}
	\caption{\small Dependence of Ne\'{e}l temperature of V$_3$Ga from disorder using calculations based on  SPR-KKR package. Disorder was obtained by exchanging a fraction x of the Ga atoms with the nonmagnetic V atoms. Note that $T_N$ at x~=~0 here is somewhat larger than that found using a more accurate simulation described in the text.}
    \label{fig5}
\end{figure}

 In order to estimate the magnetic transition temperature for the AFM G-type D0$_3$ phase, the GGA solution is more appropriate. Using Monte Carlo simulations with \textit{ab initio} exchange integrals and Heisenberg model~\cite{Supplemental}, we obtain the Ne\'{e}l temperature $T_N = 590$~K, which is somewhat higher than the peak in the experimental M(T) data at 360~K that was extracted from Fig.~\ref{fig3}. However, the Ne\'{e}l temperature is found to be strongly affected by disorder. Effects of disorder on the Ne\'{e}l temperature are calculated by using the mean field approximation implemented in the SPR-KKR packages. Figure~\ref{fig5} illustrates how $T_N$ collapses for increasing disorder when the Ga atoms are exchanged with the nonmagnetic V atoms (those V atoms lying between the two antiferromagnetically coupled magnetic V atoms). It is seen that $T_N$ goes to zero with 20\% exchange. Thus, a reduction of the Ne\'{e}l temperature to $T_N$~=~360~K would require only 6\% of the Ga atoms exchanging for nonmagnetic V atoms. However, we note that there is a sizeable energy barrier for V-Ga exchange. Nevertheless, these results illustrate the importance of the nonmagnetic V atoms to the exchange between the other two antiferromagnetically coupled V-atoms.
 
 \section{Conclusions}
 
We have studied the dual-phase superconducting A15 phase and the semiconducting AFM phase D0$_3$ in V$_3$Ga. To rationalize our results, we have considered several models within DFT.
We find that the effect of more accurate XC corrections within DFT is to eliminate or to weaken the total FM  moment given by simpler LDA and GGA schemes in the A15 cell. This FM moment could jeopardize the A15 superconducting properties. Correlation effects also tend to stabilize the D0$_3$ solution. However, this trend could be exaggerated within SCAN. Concerning the formation energies, we believe that the most accurate results are between GGA and SCAN, therefore, we deduce that A15 and D0$_3$ phases are degenerate within an uncertainty of about 10 meV/atom. Finally, assuming that GGA gives the best description for the AFM magnetic moments of V atoms in the D0$_3$ solution, we estimated the Ne\'{e}l temperature to be $T_N = 590$~K, but becomes lower with disorder in the atomic sublattices. Finally, the present results indicate the possibility of using V$_3$Ga for quantum technology devices that require both superconductivity and antiferromagnetism at the same temperature.
 
\begin{acknowledgements}
Use of the Advanced Photon Source at Argonne National Laboratory was supported by the U. S. Department of Energy, Office of Science, Office of Basic Energy Sciences, under Contract No. DE-AC02-06CH11357. The work at Northeastern University was supported by the US Department of Energy (DOE), Office of Science, Basic Energy Sciences Grant No. DE-SC0019275 (materials discovery for QIS applications) and benefited from Northeastern University’s Advanced Scientific Computation Center and the National Energy Research Scientific Computing Center through DOE Grant No. DE-AC02-05CH11231. The work of Chelyabinsk State University was supported by RSF-Russian Science Foundation project No.~17-72-20022 (computational studies). This work was supported by NSF Grants DMR-1904446 (M.E.J) and DMR-1905662 (D.H.). V.B. acknowledges support from the NUST "MISiS" No. K2-2020-018
B.B. acknowledges support from the COST Action CA16218.
\end{acknowledgements}

\label{References}
\bibliographystyle{aip} 
\bibliography{main}
\end{document}


\begin{center}

\textbf{Supplementary Material for\\ "Superconducting and Antiferromagnetic Properties of Dual-Phase V$_3$Ga"}

\end{center}

\vspace{5mm}
\textbf{Details and Results of the VASP calculations}
\vspace{5mm}

\begin{figure*}[!htb]
	\centering
	\includegraphics[width=\textwidth]{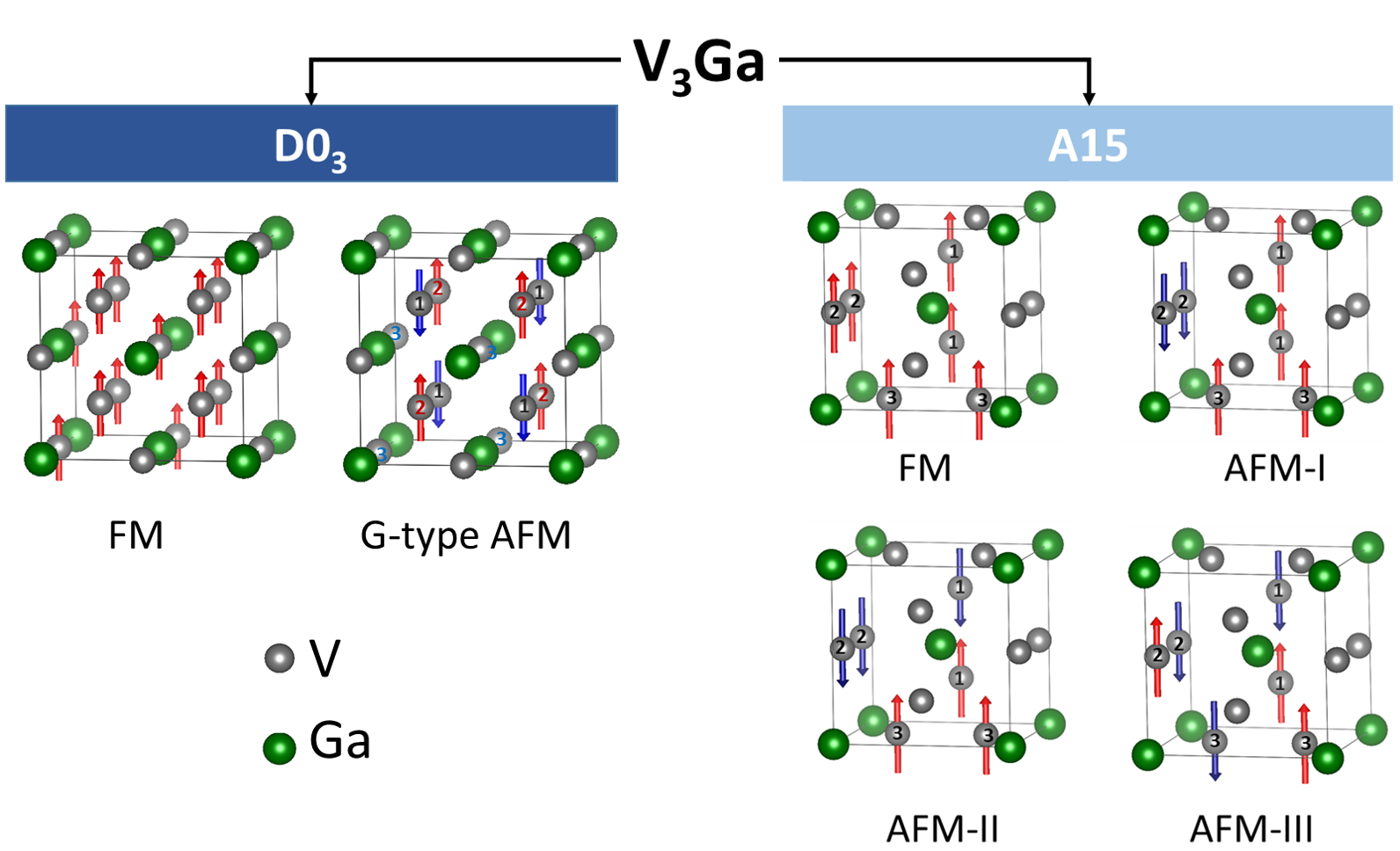}
	\caption{\small Types of the magnetic ordering for D0$_3$ and A15 structures used in calculations for V$_3$Ga.  V atoms located at different sites are noted by labels 1, 2, and 3. }
	\label{fig_1_suppl}
\end{figure*}

The geometric optimization and
ground state calculations were performed for two structures (D0$_3$ and A15) of V$_3$Ga, which are experimentally observed. 
In the case of D0$_3$ structure, the 16 atoms supercell was considered, in which Ga atoms locate at 4$a$~(0.0, 0.0, 0.0) Wyckoff positions and V atoms take 4$b$~(1/2, 1/2, 1/2) and 8$c$~((1/4, 1/4, 1/4) and 3/4, 3/4, 3/4)) sites. For A15 structure, six-atom supercell was assumed, in which two Ga atoms occupy 2$a$ sites and six V atoms locate at 6$c$ sites.
Concerning to magnetic subsystem, ferromagnetic (FM) and antiferromagnetic (AFM) spin alignments of V atoms were considered. For D0$_3$ structure, FM and AFM G-type orders were considered according to Ref.~\cite{V3AlGalanakis}, while for A15 structure, FM and three AFM orders were proposed in accordance with Ref.~\cite{He-2019}. The crystal structures as well as the magnetic configuration considered are illustrated in Fig.~\ref{fig_1_suppl}. To emphasize the AFM order, V atoms located at different sites were distinguished by labels 1, 2, and 3.

To consider the exchange-correlation effects, the following functionals were assumed in the electronic calculations: LDA-PZ~\cite{Perdew_LDA}, GGA-PBE~\cite{Perdew-1991,Burke-1997,Perdew1996}, GGA+$U$ and meta-GGA-SCAN~\cite{Perdew-1999,Tao-2003,Sun-2015} parametrizations. 
For GGA+$U$ scheme, the values of Hubbard-repulsion $U$ and Hund-exchange integral $J$ for V atoms were taken 2 and 0.67~eV, correspondingly, in accordance with  He~\textit{et al.}~\cite{He-2019}.  
The PAW pseudopotentials were taken with valence-electron configurations as follows: 3$s^2$3$p^6$4$s^2$3$d^3$ for V atoms and 3$d^{10}$4$s^2$4$p^1$ for Ga atoms. 
The 10$\times$10$\times$10 $k$-points grid was generated using Monkhorst-Pack scheme.
The kinetic-energy cutoff for the augmented charge was taken 750~eV.
The Methfessel-Paxton smearing method~\cite{Methfessel-1989} was used with a width of 0.2~eV in geometry optimization runs, and the tetrahedron smearing method with Bl\'ochl
corrections~\cite{Blochl-1994} was deployed in self-consistent calculations. The calculations were converged
to an accuracy of 10$^{-8}$ eV. The convergence criterion in the optimization for the residual forces was 10$^{-7}$ eV/\AA.

The results of \textit{ab~initio} calculations for D$0_3$ and A15 crystal structures of V$_3$Ga are presented in Tables~\ref{table-1}~and~\ref{table-2}, respectively.

\begin{table}[!htb]
	\caption{The equilibrium lattice parameter~($a_0$ in \AA), partial and total magnetic moments ($\mu_{V_i}$ and $\mu_{tot}$ in $\mu_B$/f.u.), and total energy ($E$ in eV/atom) of non-magnetic (NM) and AFM phases of D0$_3$ structure. Notice, the initial FM order was converged to the non-magnetic (NM) one. }
	\label{table-1}
		\begin{tabular}{lcccccc} 
			&
		\multicolumn{6}{c}{\begin{tabular} 
		[c]{@{}c@{}}NM\end{tabular}}    \\   \hline
			& $a_0$  & $\mu_{V_1}$ & $\mu_{V_2}$ & $\mu_{V_3}$ & \begin{tabular}[c]{@{}c@{}}$\mu_{tot}$\end{tabular} & $E$ \\ \hhline{~------}
			LDA     & 5.899 & 0.00 & 0.00 & 0.00 & 0.00 & -8.486 \\
			PBE     & 6.034 & 0.00 & 0.00 & 0.00 & 0.00 & -7.574 \\
			PBE+$U$ & 6.059 & 0.00 & 0.00 & 0.00  & 0.00 & -6.541 \\
			SCAN & 5.977 & 0.00 & 0.00       & 0.00       & 0.00                                                                  & -17.377     \\
			& \multicolumn{6}{c}{AFM}                                                                                                   \\ \hline
			& $a_0$ & $\mu_{V_1}$ & $\mu_{V_2}$ & $\mu_{V_3}$ & \begin{tabular}[c]{@{}c@{}}$\mu_{tot}$\end{tabular} & $E$ \\  \hhline{~------}
			LDA & 5.902 & -0.429 & 0.429 & 0.00 &	0.00 &	-8.487 \\
			PBE & 6.064 & -1.314 & 1.314 & 0.00 & 0.00 & -7.600        \\
			PBE+$U$ & 6.130 & -1.916 & 1.916 & 0.00  & 0.00 & -6.647 \\
			SCAN & 6.035 & -1.848 & 0.00 & 1.848 & 0.00 & -17.486  \\
		\end{tabular}%
\end{table}

\begin{table}[!!!t]
	\caption{The equilibrium lattice parameter~($a_0$ in \AA), partial and total magnetic moments ($\mu_{V_i}$ and $\mu_{tot}$ in $\mu_B$/f.u.), and total energy ($E$ in eV/atom) of FM and AFM phases of A15 structure. }
	\label{table-2}
		\begin{tabular}{lcccccc}			
			& \multicolumn{6}{c}{FM}                                                     \\ \hline
			& $a_0$       & $\mu_{V_1}$ & $\mu_{V_2}$ & $\mu_{V_3}$ & $\mu_{tot}$ & $E$       \\ \hhline{~------}
			LDA & 4.678 & 0.089 & 0.164	& 0.112 & 0.368 & -8.572 \\
			PBE        & 4.788    & 0.222       & 0.390       & 0.303       & 0.916       & -7.645  \\
			PBE+$U$     & 4.817    & 0.434       & 0.661       & 0.548       & 1.636       & -6.601  \\
			SCAN       & 4.744    & 0.308       & 0.523       & 0.414       & 1.244       & -17.442 \\
			& 
		
			 \multicolumn{6}{c}{AFM-I}                                                  \\ \hline
			& $a_0$       & $\mu_{V_1}$ & $\mu_{V_2}$ & $\mu_{V_3}$ & $\mu_{tot}$ & $E$       \\
			\hhline{~------}
			LDA & 4.678 & 0.097 & 0.122 & 0.147 & 0.368 & -8.572 \\
			PBE & 4.788 & 0.248 & 0.306 & 0.361 & 0.916 & -7.645  \\
			PBE+$U$ & 4.828 & 0.576 & 0.307 & 0.760 & 1.636 & -6.602  \\
			SCAN & 4.742 & 0.00 & 0.00 & 0.00 & 0.00 & -17.436 \\
			&  
			
			\multicolumn{6}{c}{AFM-II} \\ \hline
			& $a_0$ & $\mu_{V_1}$ & $\mu_{V_2}$ & $\mu_{V_3}$ & $\mu_{tot}$ & $E$       \\
			\hhline{~------}
			LDA & 4.677 & 0.00 & 0.00 & 0.00 & 0.00 & -8.572 \\
			PBE & 4.786 & 0.00 & 0.00 & 0.00 & 0.00  & -7.643  \\
			PBE+$U$ & 4.801 & $\pm$1.036 & $\pm$0.024 & $\pm$0.007 & 0.00 & -6.594  \\
			SCAN & 4.742 & $\pm$0.355 & $\pm$0.002   & $\pm$0.017 & 0.00 & -17.436 \\
			& 
						
			\multicolumn{6}{c}{AFM-III}  \\ \hline
			& $a_0$ & $\mu_{V_1}$ & $\mu_{V_2}$ & $\mu_{V_3}$ & $\mu_{tot}$ & $E$       \\
			\hhline{~------}
			LDA & 4.678 & 0.00 & 0.00 & 0.00 & 0.00 & -8.572 \\
			PBE & 4.788 & 0.00 & 0.00 & 0.00 & 0.00 & -7.643  \\
			PBE+$U$ & 4.879 & $\pm$1.351 & $\pm$1.508 & $\pm$1.502 & 0.00 & -6.632  \\
			SCAN & 4.744 & $\pm$0.268 & $\pm$0.326   & $\pm$0.330 & 0.00 & -17.437 \\
		\end{tabular}%
\end{table}

\vspace{5mm}
\textbf{Details and Results of the SPR-KKR calculations}
\vspace{5mm}

\begin{figure}[!htb]
	\centering
	\includegraphics[width=6cm]{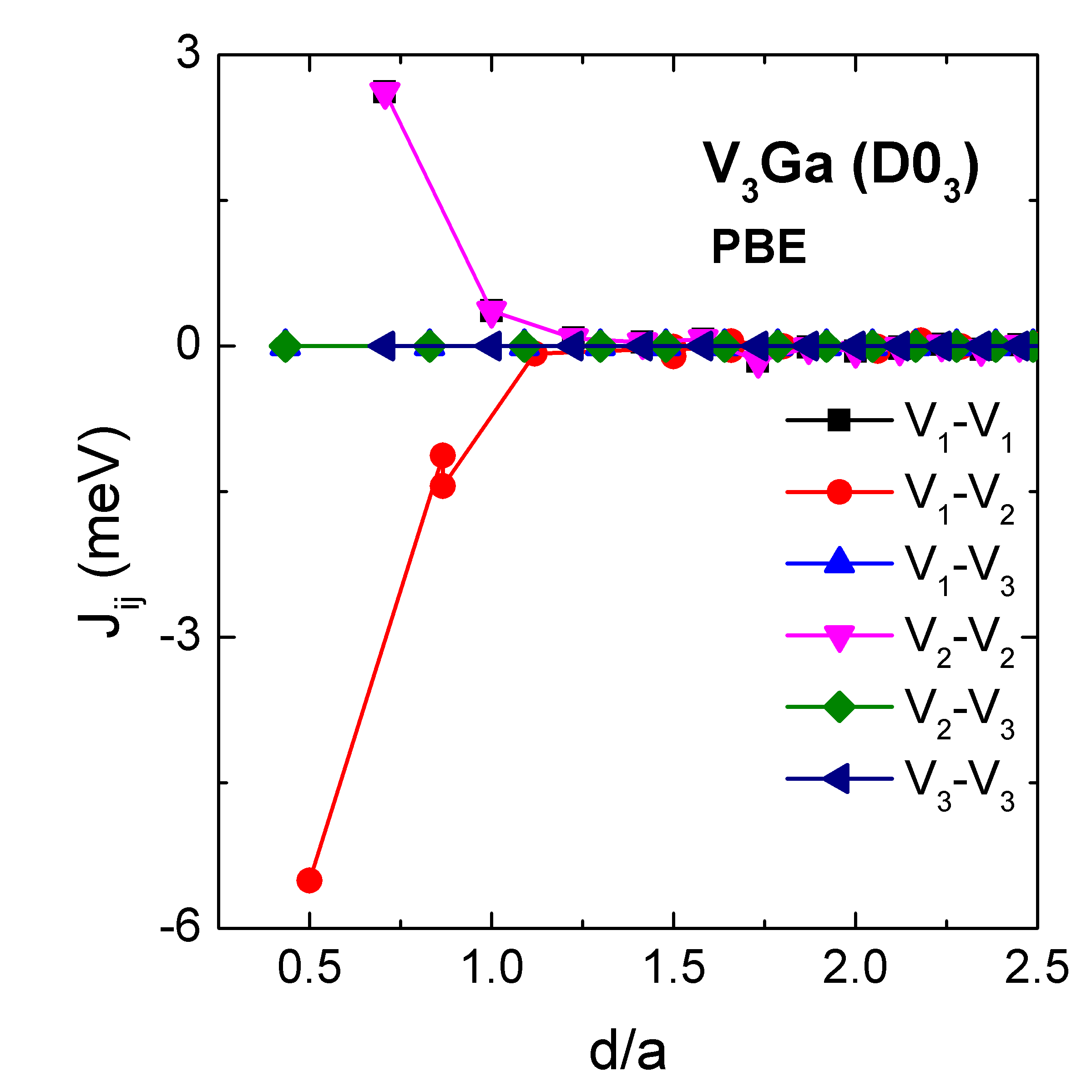}
	\caption{\small Exchange coupling constants $J_{ij}$ as a function of distance ($d/a$) between atoms $i$ and $j$ for V$_{3}$Ga with D0$_3$ structure at their equilibrium lattice parameter}
	\label{fig_Jij_suppl}
\end{figure}

The calculations of the Heisenberg's exchange coupling parameters ($J_{ij}$) were performed using of the spin-polarized
scalar-relativistic (SP-SREL) Dirac-Hamiltonian as part of the SPR-KKR (A spin-polarized relativistic Korringa-Kohn-Rostoker) package~\cite{Ebert_2011}. For these calculations, the optimized lattice parameter obtained by VASP calculation for FM phase of D0$_3$ structure has been used. It should be noted that in this case, calculations were carried out for
unit cell of D0$_3$ consisting of four atoms (3 V and 1 Ga). The exchange-correlation energy was treated by the GGA-PBE parameterization~\cite{Burke-1997}.
For the self-consistent cycles calculations, 2300 \textit{k}-points were generated by a \textit{k}-mesh grid of 45$^3$. The angular momentum expansion ($l_{max}$)
 was restricted to three.
 The magnetic exchange constants were calculated  on a \textit{k}-mesh grid of $57\times57\times57$ with 4495 \textit{k}-points. All calculations converged to 0.01 mRy of total energy.
  The dependence of the exchange interaction parameter $J_{ij}$ on the distance between atoms $d/a$ is shown in Fig.~\ref{fig_Jij_suppl}.
 
Here, the positive exchange constants ($J_{ij} > 0$) presented FM coupling, whereas the negative ones ($J_{ij} < 0$) indicate AFM coupling. It is obvious that the large FM interactions are observed between nearest neighbors V$_1$-V$_1$ and V$_2$-V$_2$ atoms. The magnitudes of these couplings are the same and about of 3 meV. While, they reduces sufficiently and becomes zero in the third and further coordination shells. The strongest AFM interaction (-$5.5$ meV) is observed between the nearest V$_1$ and V$_2$ atoms located at different sublattices. The all interactions with V$_3$ atoms are close to 0 meV during the zero magnetic moment in V$_3$ atoms (See, Table 1).

\vspace{5mm}
\textbf{Details and Results of the Monte Carlo simulations}
\vspace{5mm}

To extend the zero-temperature  \textit{ab initio} results to finite temperatures, we further performed the Monte Carlo (MC)
simulations of the classical three-dimensional Heisenberg
model ($\mathcal{H}=−\sum_{ij}J_{ij}S_{i}S_{j}$) in zero magnetic field. Here,
$\mathbf{S}_{i}=\left(S_{i}^{x},\ S_{i}^{y},\ S_{i}^{z}\right)$ is a classical Heisenberg spin variable $\left|S_i\right|=1$ and $J_{ij}$ are the exchange coupling constants (positive in case of FM interactions and negative in case of AFM interactions depending on the distance between the atoms). The calculated exchange parameters ($J_{ij}$) and partial magnetic moments ($\mu_i$) were taken as input parameters. Since, the
$J_{ij}$ behavior versus the distance between atoms ($d/a$) shows the long-range oscillation trend, therefore we restricted
exchange interactions up to sixth coordination shell for all interaction pairs.

The MC simulations were carried out for a simulation lattice with real unit cell consisting of 3925 atoms (1099 Ga and 2826 V atoms) and periodic boundary conditions by using the Metropolis algorithm~\cite{landau_binder_2005}, where, some new, random spin direction is chosen and the energy change which would result if this new spin orientation
is kept, is then calculated. Changes of the independent spin variables $\mathbf{S}_{i}=\left(S_{i}^{x},\ S_{i}^{y},\ S_{i}^{z}\right)$ are accepted or rejected according to the single-site transition probability
$W = \mathrm{min}\left[1,\exp\left(-\Delta\mathcal{H}/k_BT\right)\right]$. As  a time unit, we used one MC step consisting of $N$ attempts to change the spin variables. A new spin direction can be chosen by randomly choosing new spin components. The spin components are chosen in the following manner~\cite{landau_binder_2005}. Two random numbers $r_1$ and $r_2$ are chosen from the interval $\left[0, 1\right]$ to produce a vector with two components $\zeta_1 = 1 - 2r_1$ and $\zeta_2 = 1 - 2r_2$. The length of the vector is determined by $\zeta^2 = \zeta_1^2 + \zeta_2^2$  and if $\zeta^2<1$, then a new spin vector is computed with components
 
 $$S_i^x=2\zeta_1\sqrt{1-\zeta^2},\ S_i^y=2\zeta_2\sqrt{1-\zeta^2},\ S_i^z=1-2\zeta^2.$$

For each temperature the properties (internal energy of the system $\langle\mathcal{H} \rangle$ and magnetic order parameter $\langle m \rangle$) were estimated allowing $10^5$ MC steps and $10^4$ thermalization steps and were averaged over 2250 configurations for each 400 MC steps. The simulation started from the ferromagnetic phase with $S_i^z = 1$.

\begin{figure}[!htb]
	\centering
	\includegraphics[height=7cm]{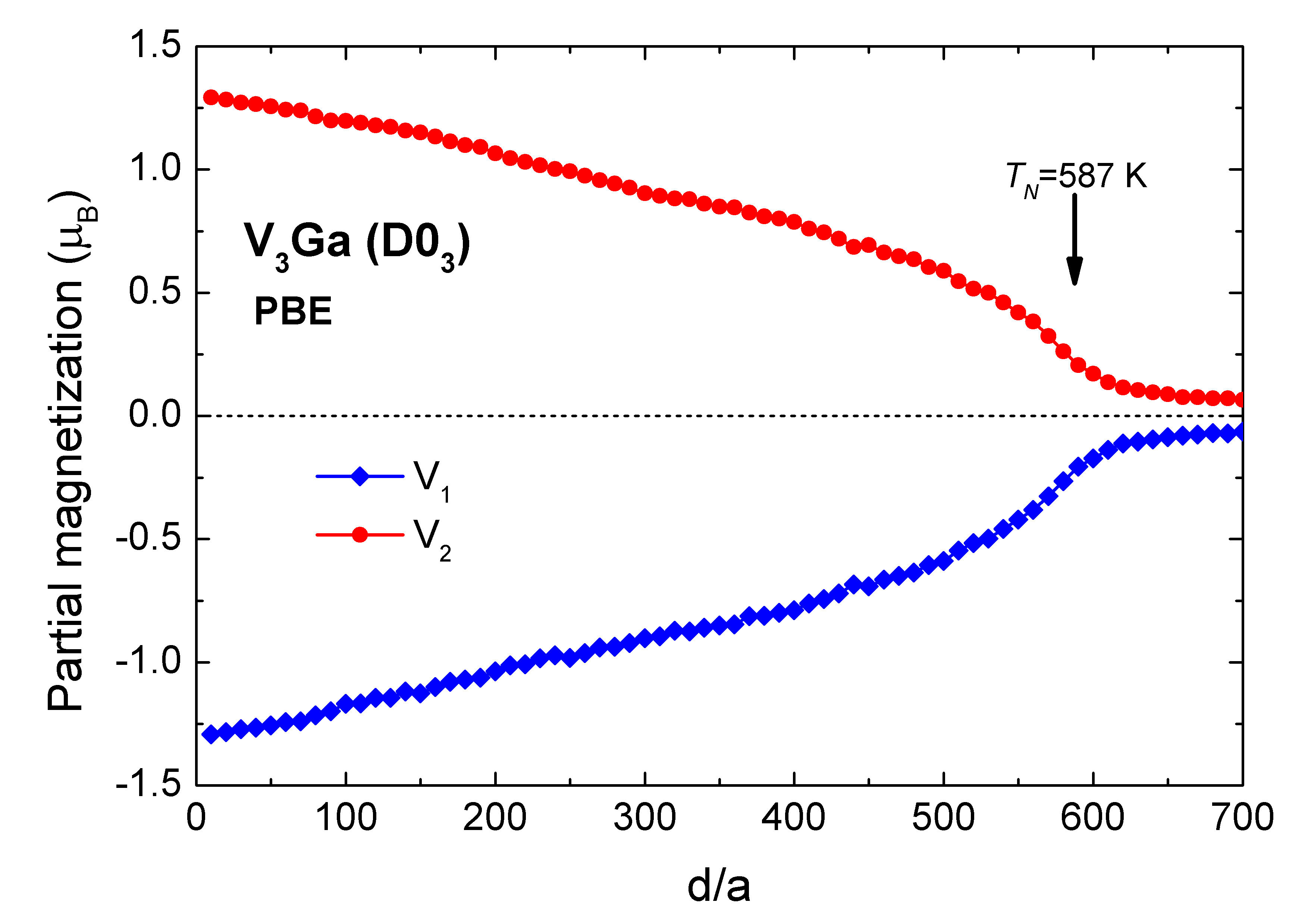}
	\caption{\small Temperature variation of the partial magnetization curves for V$_3$Ga compound with D0$_3$ structure at zero magnetic field. }
	\label{fig_MC_suppl}
\end{figure}

The magnetic order parameter is defined by the following way

$$m^{\alpha}=\frac{1}{N^{\alpha}}\sum_i\sqrt{\left( S_{i}^{\alpha,x}\right)^2+\left( S_{i}^{\alpha,y}\right)^2+\left( S_{i}^{\alpha,z}\right)^2},$$
where $\alpha$ denotes V$_1$, V$_2$ and V$_3$ atoms and $N^{\alpha}$ the total number of $\alpha$ atoms; $i$ runs over the corresponding lattice sites of the $\alpha$ atoms. The resulting temperature dependencies of partial magnetizations of V$_{1,2}$ for V$_3$Ga with D0$_3$ alloy at zero magnetic field is shown in Fig.~\ref{fig_MC_suppl}.

In order to estimate the Ne\'{e}l temperature, we plotted
$m\left(T\right) = \left(m^{\alpha}\right)^{1/\beta}\left(T\right)$ function, which approximately linear decreases with increasing temperature. The Ne\'{e}l temperature can be estimated at the intersection of $m$ with the $T$ axis. Here, $m^{\alpha}$ is the partial magnetization of V$_1$, V$_2$ atoms and $\beta$ is the critical index. Note, in the case of the three-dimensional Heisenberg model the $\beta$ exponent is equal to 0.3646. Estimation give us Ne\'{e}l temperature $T_N\ =\ 587$ K.

\vspace{5mm}
\textbf{Results of Disorder Calculations}
\vspace{5mm}

The effects of disorder on the Ne\'{e}l transition temperature was computed with SPR-KKR package and KKR-CPA approach. The degree of disorder in the atomic sublattices was quantified by $x$, given by the fraction of Ga atoms that are exchanged for nonmagnetic V atoms (labeled V$_3$), where the nonmagnetic V atoms lie between the two antiferromagnetically coupled magnet V atoms (labeled V$_1$ and V$_2$). The results of the calculations are shown in Table~\ref{table-3} for $x~=~0$ to $0.2$. For all those values of $x$, the V$_1$ and V$_2$ magnetic atoms have opposite signs and equal magnitudes, while the V$_3$ moment is zero. The Ne\'{e}l temperature is seen to decrease from $T_N\ =\ 630$ K at $x~=~0$ to $T_N\ =\ 0$ at $x~=~0.2$. Figure 5 in the main text shows $T_N$ as a function of $x$.

\begin{table}
	\caption{Partial magnetic moments of V$_i$ atoms, as well as Ne\'{e}l temperatures calculated for D0$_3$ phase of V$_3$Ga with different degree of disorder. Note, that a disorder with degree $0$ means ordered D0$_3$ phase. }
	\label{table-3}
		\begin{tabular}{lcccccc} 
			&
		    \\   \hline
			Degree of disorder & $\mu_{V_1}$  & $\mu_{V_2}$ & $\mu_{V_3}$ & $\mu_{V_4}$ & $\mu_{tot}$ & $T_N$ \\ \hhline{~------}
			0     & -1.161 & 1.161 & 0.00 & - & 0.00 & 630.0 \\
			0.05  & -0.997 & 0.997 & 0.00 & 0.00 & 0.00 & 414.5 \\
			0.1 & -0.720 & 0.720 & 0.00 & 0.00  & 0.00 & 189.3 \\
			0.2 & 0.002 & 0.002 & 0.00  & 0.00  & 0.00 & 0.00    \\
		\end{tabular}%
\end{table}

\vspace{20mm}
\textbf{Results of Rietveld X-ray Diffraction Analysis}
\vspace{5mm}

\begin{figure}[!hb]
	\centering
	\includegraphics[width = 0.95\columnwidth]{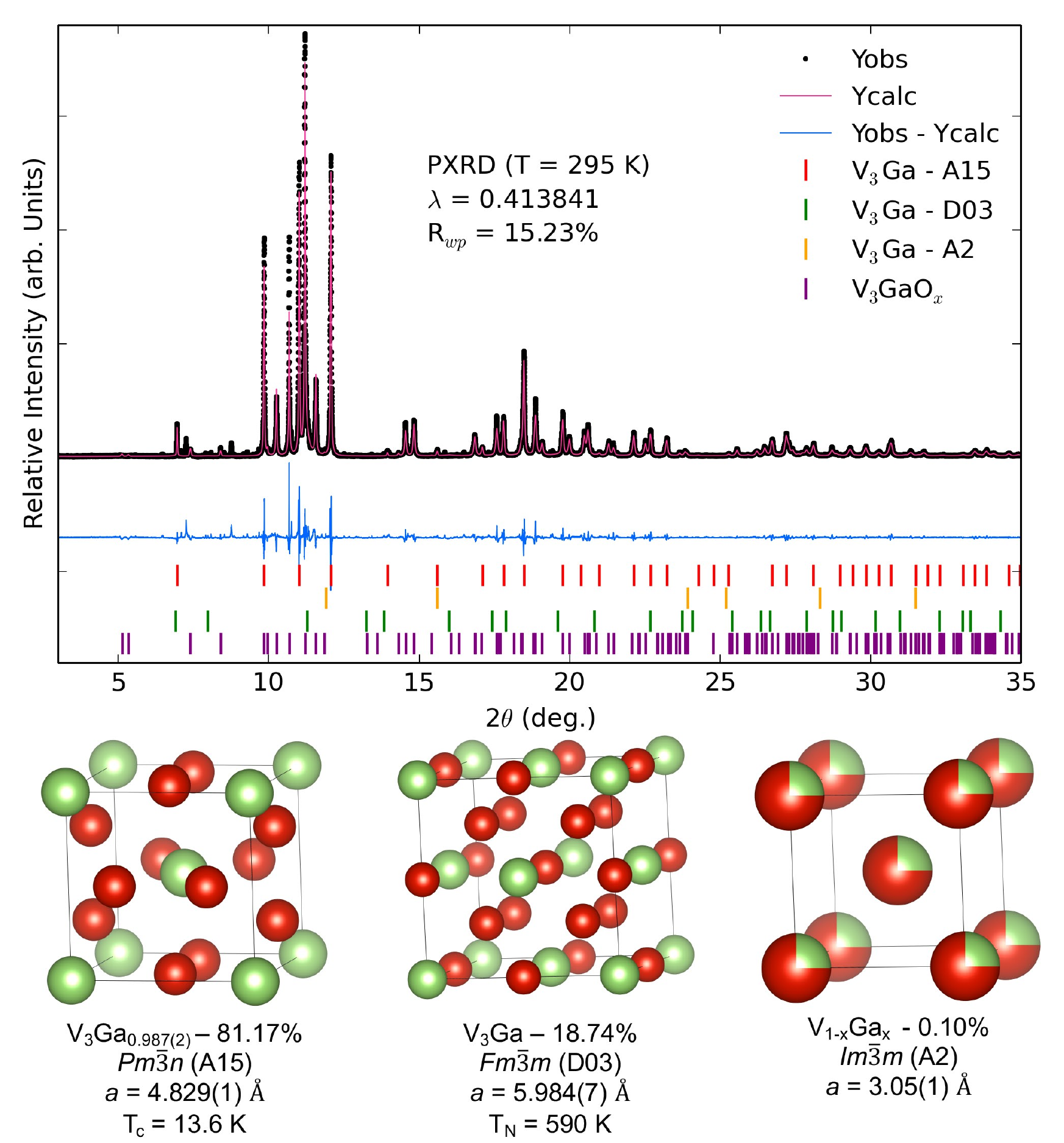}
	\caption{\small High resolution synchrotron X-ray diffraction patterns for as-grown V$_{3}$Ga ingots collected at T = 295 K. Red tick and green tick marks represent the targeted V$_{3}$Ga phases, A15 and D0$_{3}$ phases respectively. Orange tick marks represent impurity A2 phase and purple tick marks shown the existence of an impurity oxide phase, V$_{3}$GaO$_{x}$. These are shown below the calculated, observed, and differences curves from Rietveld analysis. Crystal structures, corresponding extracted lattice parameters and elemental composition from Rietveld analysis, and relevant transition temperature for each phase are shown below the XRD data.}
	\label{fig_XRD_suppl}
\end{figure}

X-ray diffraction of the V$_3$Ga sample was taken at Beamline 11-BM at the Advanced Photon Source at Argonne National Lab using a wavelength of $\lambda$ = 0.413841~\AA~ and the sample was rotated continuously. Figure~\ref{fig_XRD_suppl} shows the XRD data and subsequent Rietveld and Pawley fitting procedures. \cite{TOPAS} The experimental data is shown by the points and fit to the data is shown by the pink curve. The lower blue curve plots the experimental data minus the Rietveld fit. In this figure, only the A15, D0$_3$ and A2 phases were included as part of the Rietveld fitting procedure. The A2 phase is a disordered D0$_3$ phase where the V and Ga are randomly exchanged. Only including those three phases, it was determined that another phase was present. A search of relevant V, Ga and V-Ga binary phases (V$_{6}$Ga${5}$, V$_{6}$Ga$_{7}$, V$_{2}$Ga$_{5}$) were unable to accurately index the remaining peaks. Thus, an expanded search to contain impurity oxide phases was done. Through this, it was determined that the remaining peaks could be accurately indexed to a reported oxide impurity phase, V$_{3}$GaO$_{x}$. \cite{Maier} The structure of the V$_{3}$GaO$_{x}$ phase is not rigorously reported; however, the previous work discussed the structure as Cu$_{3}$Au-type perovskite with Ga at the corners, V at the face centers and O at the body center. From this, a structure was built and space group determined through the use of Python Materials Genomics library. \cite{Ong} Since the structure of the V$_{3}$GaO$_{x}$ has not been determined as well as its existence as a minor impurity phase in the current work, it's contribution to the X-ray diffraction pattern was modeled only by a Pawley refinement procedure. As such, the weight fraction of the V$_{3}$GaO$_{x}$ phase cannot be determined and all other reported weight fractions for the A15, D0$_{3}$, and A2 phases are reported relative to each other. The results of the fitting procedure provided the following approximate percentages of the various phases: 81\% in the A15 phase, 18\% in the D0$_3$ phase, and 0.1\% in the A2 phase. An attempt to include the determined structure for the V$_{3}$GaO$_{x}$ phase for Rietveld refinement yielded an approximate weight fraction on the order of 30 - 40\%. 

\vspace{20mm}
\textbf{Analysis of superconductivity of A15 phase}
\vspace{5mm}

\begin{figure}[!hb]
	\centering
	\includegraphics[width = 0.45\columnwidth]{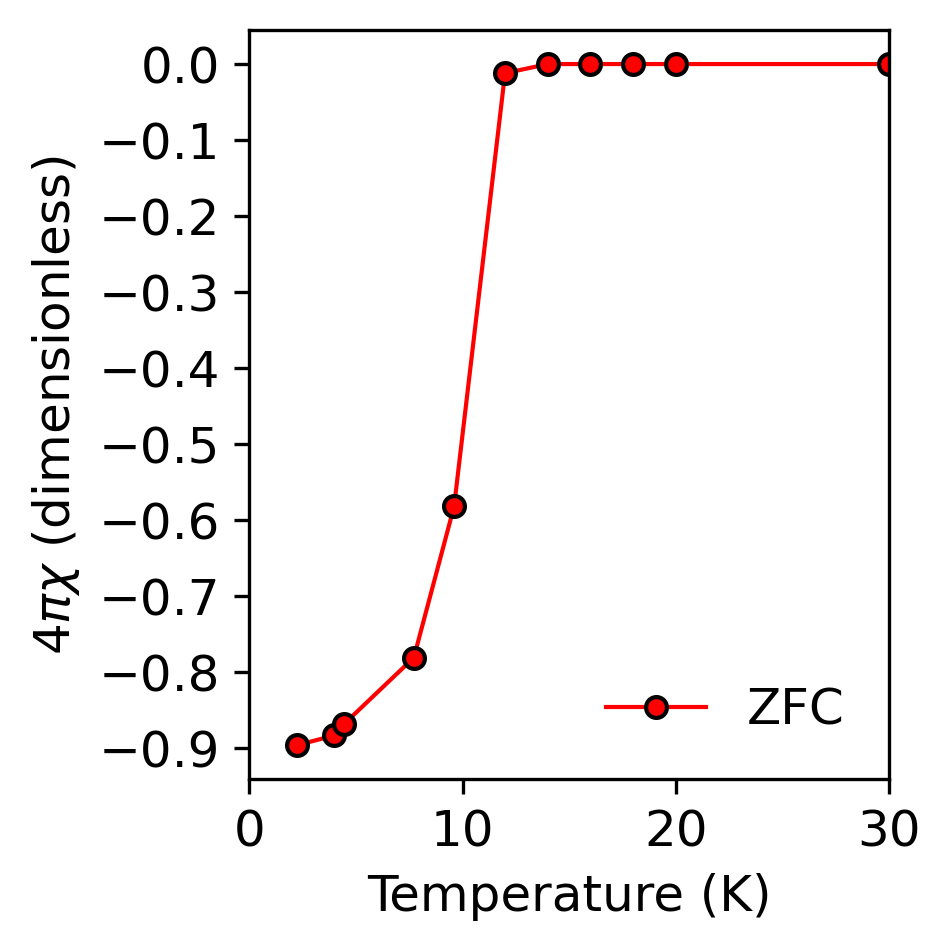}
	\includegraphics[width = 0.45\columnwidth]{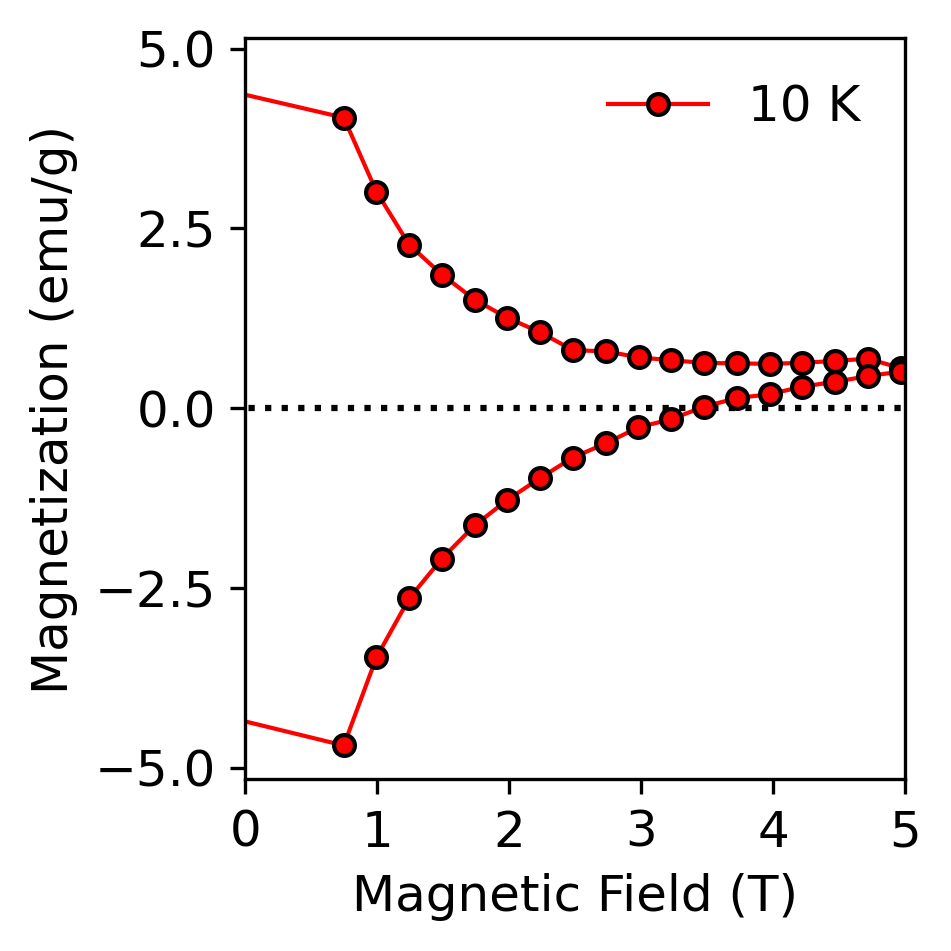}
	\caption{\small Temperature dependence of dimensionless magnetic susceptibility at 500 Oe and isothermal magnetization at 10 K of V$_{3}$Ga sample mixture measured on as-synthesized polycrystalline material.}
	\label{fig_mag_suppl}
\end{figure}

Additional analysis of the as-synthesized superconducting properties of the A15 V$_{3}$Ga revealed superconducting critical temperature of 13.6 K and a superconducting volume fraction at 2 K of approximately 90\% as shown in Fig \ref{fig_mag_suppl}. This was calculated from the density and phase fraction extracted from Rietveld analysis from synchrotron X-ray diffraction data for the A15 phase. Isothermal magnetization at 10 K revealed an upper critical field, H$_{c2}$ of about 3.5 T which is significantly lower than previous reports. \cite{Foner, Decker} This may be due to a number of factors such as stoichiometric changes in the as-synthesized A15 V$_{3}$Ga phase as well as the coexistence of another magnetic phase due to the V$_{3}$Ga D0$_{3}$ phase in the as-synthesized sample. Thus, the assignment of an H$_{c2}$ value to the A15 V$_{3}$Ga phase cannot be done in a rigorous way. However, the extracted superconducting volume fraction and superconducting critical temperature matches with expectations.

\newpage
\label{References}
\bibliographystyle{aip}
\bibliography{main2}